\documentclass{article}
\usepackage{t1enc}
\usepackage[utf8]{inputenc}
\usepackage[english]{babel}
\usepackage{graphicx}
\usepackage{amssymb}
\usepackage{amsmath}
\usepackage{amsthm}

\def\OA{{\cal A}}
\def\DK{{\cal O}}
\def\SDK{{\cal K}}
\def\PR{{\cal P}}
\def\vNA{{\cal N}}
\def\Mink{{\cal M}}
\def\CO{\mathbb{C}}
\def\RE{\mathbb{R}}
\def\IN{\mathbb{Z}}

\def\UN{\mathbf{1}}
\def\fel{{\frac{1}{2}}}

\def\qed{\ \vrule height 5pt width 5pt depth 0pt}
\def\cros{\raise1.9pt\hbox{$\scriptscriptstyle
          >$}\!\raise1.5pt\hbox{$\scriptstyle\triangleleft\,$}}
\theoremstyle{definition}\newtheorem{D}{Definition}
\theoremstyle{definition}
\theoremstyle{definition}\newtheorem{Prop}{Proposition}
\theoremstyle{definition}

\title{\bf Noncommuting local common causes for correlations violating the Clauser--Horne inequality}
\author{\textit{G\'abor Hofer-Szab\'o}\thanks{King Sigismund College, Budapest, email: gsz@szig.hu} \\
\textit{P\'eter Vecserny\'es}\thanks{Wigner Research Centre for Physics, Budapest, email: vecsernyes.peter@wigner.mta.hu}}
\date{ }

\begin{document}
\maketitle

\begin{abstract}
In the paper the EPR-Bohm scenario will be reproduced in an algebraic
quantum field theoretical setting with locally finite degrees of freedom.
It will be shown that for a set of spatially separated correlating events
(projections) maximally violating the Clauser--Horne inequality there can be given a common
causal explanation \textit{if} commutativity is abandoned
between the common cause and the correlating events. Moreover, the
noncommuting common cause will be local and supported in the common past
of the correlating events. 
\vspace{0.1in}

\end{abstract}

\section{Introduction}\label{Intro}

The central idea of algebraic quantum field theory is the representation of observables by a net of local $C^*$-algebras associated to bounded regions of a spacetime (see (Haag, 1992)). This correspondence is established \textit{via} the standard axioms of the theory such as isotony, locality (also called Einstein causality or microcausality) and covariance. A state $\phi$ in such a local quantum theory is defined as a normalized positive linear functional on the quasilocal observable algebra $\OA$ which is an inductive limit of local observable algebras. The representation $\pi_{\phi}\colon\OA\to\mathcal{B}(\mathcal{H})$ corresponding to the state $\phi$ converts the net of $C^*$-algebras into a net of von Neumann observable algebras by closures in the weak topology.

Since von Neumann algebras are rich in projections, they offer a nice representation of \textit{quantum events}: projections can be interpreted as 0-1--valued observables and their expectation value defines the probability of the event that the observable takes on the value 1 in the appropriate quantum state. Although due to the axiom of locality two projections $A$ and $B$ commute if they are contained in local algebras supported in spacelike separated regions, they can still \textit{be correlating} in a state $\phi$, that is
\begin{equation*}\label{correlation}
\phi(AB) \neq  \phi(A)\phi(B)
\end{equation*}
in general. Due to their spacelike separated supports the correlation between these events is said to be \textit{superluminal}. Since spacelike separation excludes direct causal influence, one may look for a causal explanation of these superluminal correlations in terms of \textit{common causes}.

The first probabilistic definition of the common cause is due to Hans Reichenbach (1956). Reichenbach's definition has been first generalized to algebraic quantum field theory by R\'edei (1997, 1998) and the question was posed whether superluminal correlations can be given a common causal explanation. As a positive answer, R\'edei and Summers (2002, 2007) have shown that in the von Neumann algebraic setting with type III local algebras there always exists a common cause commuting with the correlating projections in the union of their causal past called the \textit{weak past}. Hofer-Szab\'o and Vecserny\'es (2012a) have shown, however, that this is not a general fact in \textit{all} local quantum theories: there exist correlations in algebraic quantum field theory with locally \textit{finite} degrees of freedom for which no commutative common causal explanation can be given. Finally, Hofer-Szab\'o and Vecserny\'es (2012b) have shown that by allowing common causes that do \textit{not} commute with the correlating events the noncommutative version of the result of R\'edei and Summers can be regained. In the same paper it also was argued why demanding commutativity between the common causes and the correlating events is unjustified and should be given up.

This paper is a further step into the investigation of noncommuting common causes: we will apply them for the common causal explanation of not \textit{one} but \textit{several} superluminal correlations together violating the Clauser--Horne inequality. To be more specific, consider four events $A_1$, $A_2$, $B_1$ and $B_2$ such that $A_m \in \vNA(V_A)$ and $B_n \in \vNA(V_B)$ ($m,n=1,2$) where $\vNA(V_A)$ and $\vNA(V_B)$ are von Neumann algebras supported in the spacelike separated regions $V_A$ and $V_B$, respectively. Suppose that the four pairs $\{(A_m, B_n) \, \vert \, m, n =1,2\}$ are correlating in a state $\phi$ that is
\begin{equation}\label{corrs} 
\phi(A_m B_n) \neq  \phi(A_m)\phi(B_n).
\end{equation}
Suppose furthermore that they violate the Clauser--Horne (CH) inequality (Clauser and Horne, 1974) defined as follows:
\begin{eqnarray}\label{CH_int} 
-1 \leqslant \phi (A_1 B_1 + A_1 B_2 + A_2 B_1 - A_2 B_2 - A_1 - B_1 ) \leqslant 0.
\end{eqnarray}

Sometimes in the EPR-Bell literature another Bell-type inequality is used instead of (\ref{CH_int}): the Clauser--Horne--Shimony--Holte (CHSH) inequality (Clauser, Horne, Shimony and Holt, 1969) defined in the following way:
\begin{eqnarray}\label{CHSH_int} 
\left\lvert \phi \big( U_{A_1}(U_{B_1} + U_{B_2}) + U_{A_2}(U_{B_1} - U_{B_2}) \big) \right\rvert \leqslant 2
\end{eqnarray}
where $U_{A_i}$ and $U_{B_j}$ are self-adjoint \textit{contractions} (that is $-\UN \leqslant U_{A_m}, U_{B_n} \leqslant \UN$ for $m,n=1,2$) located in $\vNA(V_A)$ and $\vNA(V_B)$, respectively. It is easy to show, however, that in a given state $\phi$ the set $\{(A_m, B_n); m,n=1,2\}$ violates the CH inequality (\ref{CH_int}) \textit{if and only if} the set $\{(U_{A_m}, U_{B_n}); m,n=1,2\}$ of self-adjoint contractions given by
\begin{eqnarray}
U_{A_m} &:=& 2A_m - \UN \label{U_A} \\
U_{B_n} &:=& 2B_n - \UN \label{U_B}
\end{eqnarray}
violates the CHSH inequality (\ref{CHSH_int}). Therefore we will concentrate only on the CH-type Bell inequalities.

A number of important results in (Summers, Werner 1988), (Halvorson, Clifton 2000) prove that Poincar\'e covariant algebraic field theories abound in normal states establishing superluminal correlations between spacelike separated projections/contractions violating the CH/CHSH inequality. Here we produce such states in algebraic field theories with locally finite degrees of freedom, namely in local UHF-type quantum theories (Hofer-Szab\'o and Vecserny\'es, 2012b). In this case the events $A_n$ and $B_m$ will be elements of such commuting local algebras $\vNA(V_A)$ and $\vNA(V_B)$ with spacelike separated supports $V_A$ and $V_B$ that are isomorphic to simple matrix algebras $M_A$ and $M_B$. Moreover, they generate a tensor product algebra: $\vNA(V_A)\vee\vNA(V_B)\simeq M_A\otimes M_B$. Thus the meaning of the mentioned inequalities can be enlightened in terms of quantum information theory (see for example (Horodeczki et al., 2009)): The existence of a correlation (\ref{correlation}) simply means that $\phi$ is not a product state on $M_A\otimes M_B$. Correlations can be obtained from separable states, i.e. from convex combination of product states. Since a simple matrix algebra admits a unique normalized trace, states can be  uniquely given by density matrices. Hence, product states are characterized by tensor product density matrices and separable states are characterized by separable density matrices, i.e. by convex combinations of tensor product density matrices. Since in the real linear space of selfadjoint elements in $M_A\otimes M_B$ positive elements $(M_A\otimes M_B)_+$ form a self-dual cone with respect to the scalar product $(H_1,H_2):={Tr}(H_1H_2)$, the (convex) set of states, that is the (convex) set of density matrices is even more rich: there also exist non-separable that is entangled states on $M_A\otimes M_B$. This is due to the existence of positive elements in $M_A\otimes M_B$ that cannot be written as positive linear combinations of tensor product of positive elements in $M_A$ and $M_B$. Entangled states lead to inherent and numerically stronger correlations among events contained in different tensor factors of $M_A\otimes M_B$. Separable and entangled states can be distinguished by witness operators $w\in\hat S\setminus (M_A\otimes M_B)_+$ which are the non-positive elements in the dual cone $\hat S$ of the cone $S$ spanned by separable density matrices. Clearly, a witness operator $w$ gives rise to inequalities since $\phi(w)\geq 0$ for any separable state $\phi$ on $M_A\otimes M_B$. Basically, these properties are behind the inequalities for superluminal correlations discussed above. 

Coming back to the superluminal correlations (\ref{corrs}) violating the CH inequality (\ref{CH_int})/CHSH inequality (\ref{CHSH_int}), it is well known that the violation of these inequalities leads to various no-go theorems excuding a local common causal explanation of the set (see for example (Shimony 2004) and the literature therein). However, all these common causal explanations assume that the common cause is \textit{commuting} with the correlating events. Hence, a natural question arises: Does the noncommutative generalization of the common causes in algebraic quantum field theory give enough freedom to find \textit{noncommuting} common causes for a set of correlations violating the CH/CHSH inequality?

In this paper we will answer this question in the affirmative by showing that the violation of the CH/CHSH inequality does \textit{not} exclude the existence of a common causal explanation of the set of correlations in question \textit{if} we abandon commutativity between the common cause and the outcome events. Moreover, in the given example the noncommuting common cause turns out to be local and supported in the common past i.e. in the intersection of the causal pasts of the correlating events.

In Section 2 we reproduce the EPR-Bohm scenario in the local quantum Ising model, which is a prototype of local UHF-type quantum theories. Giving a faithful state on the corresponding observable algebra we show that there is a set $\{(A_m, B_n); m,n=1,2\}$ of projections having such correlations that maximally violates the CH inequality. In Section 3 we introduce the main concepts of a common causal explanation of superluminal correlations and we explicitly construct a common cause of the above four correlations. We conclude in Section 4.

\section{Correlations violating the Clauser--Horne inequality}\label{sec:Correlations}

\begin{figure}[ht]
\centerline{\resizebox{6cm}{!}{\includegraphics{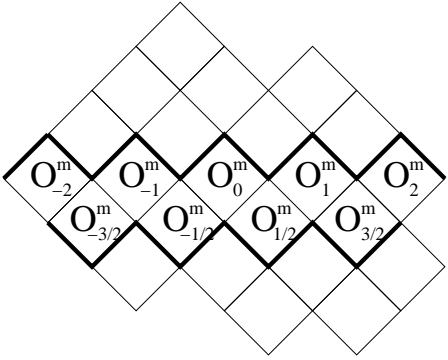}}}
\caption{A thickened Cauchy surface in the two dimensional Minkowski space $\Mink^2$}
\label{strong_Bell1}
\end{figure}

Consider the net of `intervals' $(i,j):=\{i,i+\fel,\dots,j-\fel,j\}\subset\fel\mathbb{Z}$ of half-integers. The set of half-integers can be interpreted as the space coordinates of the center $(0,x),x\in\mathbb{Z}$ and $(-1/2,x),x\in\mathbb{Z}+1/2$ of minimal double cones $\DK^m_x$ of unit diameter on a 'thickened' Cauchy surface in two dimensional Minkowski space $M^2$. (See Fig. \ref{strong_Bell1}.) An interval $(i,j)\subset\frac{1}{2}\mathbb{Z}$ can be interpreted as the smallest double cone $\DK_{i,j}:=\DK_i^m\vee\DK_j^m\subset M^2$ containing both $\DK_i^m$ and $\DK_j^m$. They determine a directed subset $\SDK^m_{CS}$ of double cones in $\Mink^2$, which is left invariant by the group of space-translations with integer values. 
The net of local algebras is defined as follows. The `one-point' observable algebras $\OA(\DK^m_i),\ i\in\fel\mathbb{Z}$ associated to the minimal double cone $\DK^m_i$ are defined to be isomorphic to the direct sum matrix algebra $M_1(\CO)\oplus M_1(\CO)$. 
Let $U_i$ denote the selfadjoint unitary generator of the algebra $\OA(\DK^m_i), i\in\fel\mathbb{Z}$. Between the generators one demands the following commutation relations:
\begin{eqnarray}\label{comm_rel}
U_i U_j = \left\{ \begin{array}{rl} -U_j U_i, & \mbox{if}\ |i-j|=\fel
\\ U_j U_i, & \mbox{otherwise.}\ \end{array} \right.
\end{eqnarray}
Now, the local algebras $\OA(\DK_{i,j}), \DK_{i,j}\in\SDK^m_{CS}$ are defined to be the linear span of the monoms
\begin{eqnarray}\label{monoms}
U_i^{k_i} \, U_{i+\fel}^{k_{i+\fel}} \,  \dots \, U_{j-\fel}^{k_{j-\fel}} \, U_j^{k_j} 
\end{eqnarray}
where $k_i, k_{i+\fel} \dots k_{j-\fel}, k_j \in \{0,1\}$.\footnote{For detailed Hopf algebraic description of the local quantum spin models see (Szlach\'anyi, Vecserny\'es, 1993), (Nill, Szlach\'anyi, 1997), (M\"uller, Vecserny\'es)).} 

Specifically, the 'two-point' algebra $\OA(\DK_{j,j+\fel})\simeq M_2(\CO)\,\,   (j\in\frac{1}{2}\mathbb{Z})$ is spanned by the selfadjoint unitary monoms 
\begin{eqnarray}\label{monoms2}
\UN,\qquad U_j,\qquad U_{j+\fel},\qquad iU_j U_{j+\fel},
\end{eqnarray}
(where $i$ is the imaginary unit). They satisfy the same commutation relations like the Pauli matrices $\sigma_0 = \UN, \sigma_x, \sigma_y$ and $\sigma_z$ in $M_2(\CO)$. The minimal projections in $\OA(\DK_{j,j+\fel})$ can be parametrized by a unit vector ${\bf r} = (r_1, r_2, r_3)$ in $\RE^3$ as follows:
\begin{eqnarray}
P=\fel\left( \UN+r_1 U_j+r_2 U_{j+\fel} + r_3 iU_j U_{j+\fel}\right) 
\end{eqnarray}
and can be interpreted as the event localized in $\DK_{j,j+\fel}$ pertaining to the generalized spin measurement in direction ${\bf r}$.

Similarly, the 'three-point' algebra $\OA(\DK_{j-\fel,j+\fel})\simeq M_2(\CO)\oplus M_2(\CO)\, \, (j\in\frac{1}{2}\mathbb{Z})$ is linearly spanned by the selfadjount unitary monoms 
\begin{eqnarray}\label{monoms3}
\UN,\quad U_{j-\fel},\quad U_j,\quad U_{j+\fel},\quad iU_{j-\fel} U_j,\quad iU_j U_{j+\fel}, \quad U_{j-\fel} U_{j+\fel},\quad U_{j-\fel} U_j U_{j+\fel}, 
\end{eqnarray}
and any minimal projection in $\OA(\DK_{j-\fel,j+\fel})$ can be written in the form 
\begin{eqnarray}
P=\frac{1}{4}\left(\UN \pm U_{j-\fel} U_{j+\fel}\right)\left(\UN+r_1 U_j+r_2 U_{j+\fel} + 
r_3 iU_j U_{j+\fel}\right)
\end{eqnarray}
where ${\bf r}$ is again a unit vector in $\RE^3$.

Since the local algebras $\OA(\DK_{j,j-\fel+n}), \, i\in\fel\mathbb{Z}$ for $n\in \mathbb{N}$ are isomorphic to the full matrix algebra $M_{2^n}(\mathbb{C})$ the quasilocal observable algebra $\OA$ is a uniformly hyperfinite (UHF) $C^*$-algebra, which possesses a unique (non-degenerate) normalized trace $\textrm{Tr}\colon\OA\to\CO$.

Now, consider the subset $\SDK^m$ of double cones in $\Mink^2$ that are spanned by minimal double cones being integer time translates of those in $\SDK^m_{CS}$. The directed set $(\SDK^m,\subseteq )$ is left invariant by integer space and time translations. In order to extend the ''Cauchy surface net'' $\{\OA(\DK),\DK\in\SDK^m_{CS}\}$ to the net $\{\OA(\DK),\DK\in\SDK^m\}$ in a causal and time translation covariant manner one has to classify causal (integer valued) time evolutions in the local quantum Ising model. This classification was given in (M\"uller, Vecserny\'es) and it also was shown that the extended net satisfies isotony, Einstein causality, algebraic Haag duality
\begin{eqnarray}\label{Haag} 
\OA(\DK')'\cap\OA =\OA(\DK),\quad\DK\in\SDK^m,
\end{eqnarray}
and primitive causality:
\begin{eqnarray}\label{primcaus} 
\OA(V)=\OA(V'').
\end{eqnarray}
Here $V$ is a finite connected piece of a thickened Cauchy surface (composed of minimal double cones) and $V''$ denotes the double spacelike complement of $V$, which is the smallest double cone in $\SDK^m$ containing $V$.\footnote{Since $V\notin\SDK$ we note that $\OA(V)$ is defined to be the $C^*$-algebra in $\OA$ generated by the algebras $\OA(\DK^m), \DK^m\subset V$.} Moreover, the commuting (unit) time and (unit) space translation automorphisms $\beta$ and $\alpha$ of the quasilocal algebra $\OA$ act covariantly on the local algebras, i.e. $\{\OA(\DK),\DK\in\SDK^m\}$ is a $\IN\times\IN$-covariant local quantum theory. The causal time translation automorphisms $\beta$ of $\OA$ can be parametrized by $\theta_1,\theta_2;\eta_1,\eta_2$ with $-\pi/2 <\theta_1,\theta_2\leq\pi/2$ and $\eta_1,\eta_2\in\{1,-1\}$ and they are given on the algebraic generator set $\{U_i\in\OA(\DK^m_i), i\in\frac{1}{2}\mathbb{Z}\}$ of $\mathcal A$. The automorphisms $\beta=\beta(\theta_1,\theta_2,
\eta_1,\eta_2)$ of $\mathcal A$ corresponding to causal time translations by a unit read as
\begin{eqnarray}\label{causal_automorph1}
\beta(U_x)&=&\eta_1 \sin^2\theta_1 U_x+\eta_1 
      \cos^2\theta_1 \beta(U_{x-\fel})U_x\beta(U_{x+\fel})\nonumber\\
  &&\quad+i\sin\theta_1\cos\theta_1(\beta(U_{x-\fel})U_x-U_x\beta(U_{x+\fel})),\\
\label{causal_automorph2}
\beta(U_{x+\fel})&=&\eta_2\sin^2\theta_2 U_{x+\fel} 
         +\eta_2\cos^2\theta_2U_xU_{x+\fel}U_{x+1}\nonumber\\
 &&\quad+i\sin\theta_2\cos\theta_2(U_xU_{x+\fel}-U_{x+\fel}U_{x+1}),
\end{eqnarray}
where $x\in\IN$.

It was also shown (see M\"uller, Vecserny\'es) that the following algebra isomorphisms hold: If $\DK\in\SDK^m$ is a double cone containing $n_+$ and $n_-$ minimal double cones in the right forward and left forward lightlike directions, respectively, then $\vert\OA(\DK)\vert$, the linear dimension of the corresponding local algebra is $2^{n(\DK)}, n(\DK):=n_++n_--1$ and
\begin{eqnarray}\label{localgtype}
\OA(\DK) \simeq \left\{ \begin{array}{rl} 
M_{2^{n(\DK)/2}}(\CO), & \mbox{if}\ n(\DK)\ \mbox{is even,}\\ 
M_{2^{(n(\DK)-1)/2}}(\CO)\oplus M_{2^{(n(\DK)-1)/2}}(\CO), 
                       & \mbox{if}\ n(\DK)\ \mbox{is odd.} \end{array} \right.
\end{eqnarray}

\begin{figure}[ht]
\centerline{\resizebox{6cm}{!}{\includegraphics{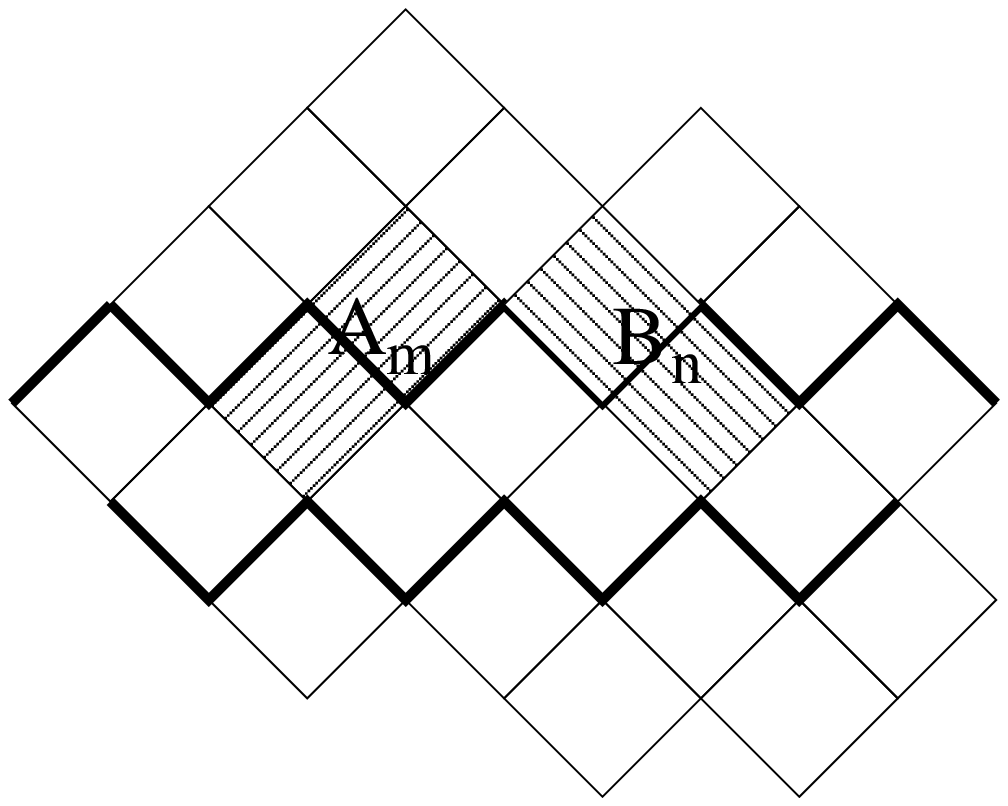}}}
\caption{Projections in $\OA(\DK_A)$ and $\OA(\DK_B)$}
\label{strong_Bell2}
\end{figure}

After this general introduction let us specify our model. Let $\DK^m(t,i)$ denote the minimal double cone with time $t$ and space $i$ coordinates of its center. Consider the double cones $\DK_A:=\DK^m(0,-1)\cup\DK^m(1,-\fel)$ and $\DK_B:=\DK^m(1,\fel)\cup\DK^m(0,1)$ and the 'two-point' algebras  $\OA(\DK_A)$ and $\OA(\DK_B)$ pertaining to them. Since they do \textit{not} lie in the original Cauchy surface (see Fig. \ref{strong_Bell2}), 
where the generators $U_i,i\in\fel\IN$ are supported, we have to use a causal time evolution $\beta$ to give elements of them. Let $A_m; m=1,2$ and $B_n; n=1,2$ be minimal projections in $\OA(\DK_A)$ and in $\OA(\DK_B)$, respectively, given by
\begin{eqnarray}
A_m &=&\fel\left( \UN+a^m_1 U_{-1} + a^m_2 \beta(U_{-\fel}) + a^m_3 iU_{-1} \beta(U_{-\fel})\right)  \label{A_m} \\
B_n &=&\fel\left( \UN+b^n_1 U_1 - b^n_2 \beta(U_{\fel}) + b^n_3 i\beta(U_{\fel}) U_{1}\right)  \label{B_n} 
\end{eqnarray}
where ${\bf a^m} = (a^m_1, a^m_2, a^m_3)$ and ${\bf b^n} = (b^n_1, b^n_2, b^n_3)$ are four ($m,n = 1,2$) unit vectors in $\RE^3$. 

The state $\phi^s$ on $\OA$ will be given by a density matrix $\rho^s$ using the normalized trace on $\OA$: $\phi^s(-):=Tr(\rho^s -)$. The density matrix $\rho^s$ is chosen in $\OA(\DK_A)\vee\OA(\DK_B)\simeq M_2(\CO)\otimes M_2(\CO)$ to describe the usual `singlet state' by restricting $\phi^s$ to this subalgebra of $\OA$. That is in terms of matrix units $e_{ij}\in M_2(\CO); i,j=1,2$: 
\begin{eqnarray} \label{rho1} 
\rho^s &:=& 2 \left( e^L_{11} \otimes e^R_{11} + e^L_{22} \otimes e^R_{22} - e^L_{12} \otimes e^R_{12} - e^L_{21} \otimes e^R_{21} \right) .
\end{eqnarray}
In terms of local generators the matrix units are given as 
\begin{eqnarray}
e^{L/R}_{11} &=& \frac{1}{2} \big(\UN + \beta(U_{-\fel/\fel}\big) \label{P1}, \\
e^{L/R}_{22} &=& \frac{1}{2} \big(\UN - \beta(U_{-\fel/\fel}\big), \\
e^{L/R}_{12} &=& \frac{1}{2} \big(\UN + \beta(U_{-\fel/\fel}\big)U_{-1/1}, \\
e^{L/R}_{21} &=& \frac{1}{2} \big(\UN - \beta(U_{-\fel/\fel}\big)U_{-1/1}, \label{P4} 
\end{eqnarray}
hence, substituting (\ref{P1})-(\ref{P4}) in (\ref{rho1}) one gets
\begin{eqnarray} \label{rho2} 
\rho^s &=& \UN + \beta(U_{-\fel}) \beta(U_{\fel}) - U_{-1} U_1 + U_{-1} \beta(U_{-\fel}) \beta(U_{\fel}) U_1. 
\end{eqnarray}
Since $\rho^s$ is a scalar multiple of a minimal projection in $\OA(\DK_A)\vee\OA(\DK_B)$, the state $\phi^s$ is not locally faithful. To get a locally faithful state one can use the convex combination
\begin{eqnarray} \label{rho3}
\rho = \lambda \, \rho^s + (1-\lambda) \, \UN
\end{eqnarray}
which gives back the singlet state for $\lambda = 1$.

The correlation between $A_m$ and $B_n$ in the state $\phi(-):=Tr(\rho-)$ is defined as the deviation from the product state value:
\begin{eqnarray}\label{corr} 
corr(A_m,B_n) &:=& \phi(A_m B_n) - \phi(A_m)\phi(B_n)\\
               &=& \phi(A_mB_n)\phi(A_m^{\perp}B_n^{\perp}) 
                  -\phi(A_mB_n^{\perp})\phi(A_m^{\perp}B_n).\nonumber
\end{eqnarray}
Using (\ref{A_m})-(\ref{B_n}) and (\ref{rho3}) this correlation reads as
\begin{eqnarray}\label{corr'} 
corr(A_m,B_n) = - \frac{\lambda}{4} \left\langle {\bf a^m}, {\bf b^n}\right\rangle 
\end{eqnarray}
where $\left\langle\ ,\ \right\rangle$ is the scalar product in $\RE^3$. Therefore $A_m$ and $B_n$ will correlate whenever ${\bf a^m}$ and ${\bf b^n}$ are not orthogonal.
Now, let ${\bf a^m}$ and ${\bf b^n}$ be chosen as follows:
\begin{eqnarray}
{\bf a^1} &=& (0,1,0) \label{a1} \\
{\bf a^2} &=& (1,0,0) \\
{\bf b^1} &=& \frac{1}{\sqrt{2}} (1,1,0) \\
{\bf b^2} &=& \frac{1}{\sqrt{2}} (-1,1,0) \label{b2} 
\end{eqnarray}
For this setting the Clauser--Horne inequality (\ref{CH_int}) will be violated at the lower bound since
\begin{eqnarray}\label{CH'} 
\phi( A_1 B_1 + A_1 B_2 + A_2 B_1 - A_2 B_2 - A_1 - B_1)= \nonumber\\
-\fel - \frac{\lambda}{4} \left(  \left\langle {\bf a^1}, {\bf b^1} \right\rangle + \left\langle {\bf a^1}, {\bf b^2} \right\rangle + \left\langle {\bf a^2}, {\bf b^1} \right\rangle - \left\langle {\bf a^2}, {\bf b^2} \right\rangle \right) = - \frac{1+\lambda\sqrt{2}}{2} ,
\end{eqnarray}
which is smaller than $-1$ if $\lambda > \frac{1}{\sqrt{2}}$. Or, equivalently, the Clauser--Horne--Shimony--Holte inequality (\ref{CHSH_int}) (with $U_{A_m}$ and $U_{B_n}$ defined in (\ref{U_A})-(\ref{U_B})) will be violated for $\lambda > \frac{1}{\sqrt{2}}$ since then
\begin{eqnarray}\label{CHSH'} 
\phi(U_{A_1}(U_{B_1} + U_{B_2}) + U_{A_2}(U_{B_1} - U_{B_2}))= \nonumber \\
=  -\lambda\left( \left\langle {\bf a^1}, {\bf b^1} + {\bf b^2}\right\rangle + \left\langle {\bf a^2}, {\bf b^1} - {\bf b^2}\right\rangle  \right) = -\lambda 2 \sqrt{2}< -2 .
\end{eqnarray}
Both the CH and the CHSH inequality are maximally violated if $\lambda = 1$ that is for the `singlet state', which is a maximally entangled state for $M_2(\CO)\otimes M_2(\CO)$.

Then the question arises: Does the above set $\{(A_m, B_n); m,n=1,2\}$ violating the CH inequality have a common causal explanation? To answer this question we have to characterize what a common causal explanation consists in.

\section{Noncommuting joint common causes for correlations violating the Clauser--Horne inequality}

As mentioned in the Introduction, the first characterization of the common cause in terms of a classical probability measure space is due to Reichenbach (1956). Although Reichenbach's original defintion is inappropriate to explain correlations in algebraic quantum field theory for several reasons (it is classical, it does not take into account the spacetime localization of the events, etc.; for details see (Hofer-Szab\'o and Vecserny\'es, 2012a)), one can easily generalize the notion of the common cause for the non-classical case. Here we repeat only the definition given in (Hofer-Szab\'o and Vecserny\'es, 2012a) because it also embraces the commuting common causes defined in (R\'edei 1997, 1998).

Let $\PR(\vNA)$ be the (non-distributive) lattice of projections (events) in a von Neumann algebra $\vNA$ and let $\phi\colon\vNA\to\CO$ be a state on it. A set of mutually orthogonal projections $\left\{ C_k \right\}_{k\in K}\subset\mathcal{P}(\vNA)$ is called a \textit{partition of the unit} $\UN\in\vNA$ if $\sum_k C_k = \UN$. Such a partition defines a
\textit{conditional expectation}
\begin{equation}\label{ncqcorr}
E\colon\vNA\to{\cal{C}}, \, \, A\mapsto 
  E(A):=\sum_{k\in K} C_kAC_k,
\end{equation}
that is $E$ is a unit preserving positive surjection onto the unital
$C^*$-subalgebra  ${\cal{C}}\subseteq\vNA$ obeying the bimodule property
$E(B_1AB_2)=B_1E(A)B_2; A\in\vNA, B_1, B_2\in{\cal{C}}$. We note that
${\cal{C}}$ contains exactly those elements of $\vNA$ that commute with
$C_k,k\in K$. Since $\phi\circ E$ is also a state on $\mathcal{N}$ we can give the following  

\begin{D}\label{ncqccs}
A partition of the unit $\left\{ C_k \right\}_{k\in
K}\subset\mathcal{P}(\mathcal{N})$ is said to be a {\em common cause system} of the commuting events $A,B\in\cal{P}(\vNA)$, which correlate in the state
$\phi\colon\vNA\to\CO$, if 
\begin{eqnarray}\label{ncqccs1}
\frac{(\phi\circ E)(ABC_k)}{\phi(C_k)}&=& \frac{(\phi\circ E)(AC_k)}{\phi(C_k)}
\frac{(\phi\circ E)(BC_k)}{\phi(C_k)}
\end{eqnarray} 
for $k\in K$ with $\phi(C_k)\not= 0$. 
If $C_k$ commutes with both $A$ and $B$ for all $k\in K$ we call $\left\{ C_k
\right\}_{k\in K}$ a {\em commuting} common cause system, otherwise a {\em noncommuting} one. A common cause system of size $\vert K\vert=2$ is called a {\em common cause}. 
\end{D}

Some remarks are in place here. First, in case of a commuting common cause
system $\phi\circ E$ can be replaced by $\phi$ in (\ref{ncqccs1}) since
$(\phi\circ E)(ABC_k)=\phi(ABC_k), k\in K$. Second, using the decompositions of
the unit, $\UN=A+A^{\perp}=B+B^{\perp}$, (\ref{ncqccs1}) can be rewritten in an
equivalent form:
\begin{equation}\label{ncqccsrew}
(\phi\circ E)(ABC_k))(\phi\circ E)(A^{\perp}B^{\perp}C_k)
=(\phi\circ E)(AB^{\perp}C_k)(\phi\circ E)(A^{\perp}BC_k),\ k\in K.
\end{equation}
One can even allow here the case $\phi(C_k)=0$ since then both sides of
(\ref{ncqccsrew}) are zero. 

We also have to specify the spacetime localization of the possible common causes of the correlations. One can define different pasts of the bounded contractible regions $V_1$ and $V_2$ in a spacetime $\mathcal{S}$ as:
\begin{eqnarray*}\label{wcspast}
\makebox{\textit{weak past:}} &&  wpast(V_1, V_2) := I_-(V_1)\cup I_-(V_2) \\
\makebox{\textit{common past:}} && cpast(V_1, V_2) := I_-(V_1)\cap I_-(V_2) \\
\makebox{\textit{strong past:}} && spast(V_1, V_2) := \cap_{x \in V_1 \cup V_2}\, I_-(x)
\end{eqnarray*}
where $I_-(V)$ denotes the union of the backward light cones $I_-(x)$ of every point $x$ in $V$ (R\'edei, Summers 2007). Clearly, ${wpast}\supset {cpast}\supset {spast}$.

With these different localizations of the common cause in hand now we can define various
common cause systems in local quantum theories according to (i) whether commutativity is required and (ii) where the common cause system is localized. Thus we can speak about \textit{commuting/noncommuting (weak/strong) common cause systems}. 

Since in our model of the previous Section we have not one pair of correlating events but four, we need also the notion of the joint\footnote{In (Hofer-Szab\'o and Vecserny\'es, 2012a,b) called \textit{common} common cause system.} common cause system: 
\begin{D}\label{lcccs}
Let $\{\OA(V),V\in\SDK\}$ be the observable net in a local quantum theory and let $\{A_m;m = 1,\dots ,M\}$ and $\{B_n;n = 1,\dots ,N\}$ be finite sets of projections in the algebras $\OA(V_1)$ and $\OA(V_2)$, respectively, supported in spacelike separated regions $V_1,V_2\in\SDK$. Suppose that all pair of spacelike separated projections $(A_m,B_n)$ correlate in a state $\phi$ of $\OA$. Then the set of correlations is said to possess a commuting/noncommuting (weak/strong) \textit{joint} common cause system if there exists \textit{a single} commuting/noncommuting (weak/strong) common cause system for \textit{all} correlations $(A_m, B_n)$.
\end{D}
\vspace{0.1in}

\noindent
Having defined what a common causal explanation means we answer the question raised at the end of the previous Section in the affirmative: we explicitly contruct a noncommuting joint common cause for the set $\{(A_m, B_n); m,n=1,2\}$ of correlations violating the CH inequalities (\ref{CH_int}).

\begin{figure}[ht]
\centerline{\resizebox{7cm}{!}{\includegraphics{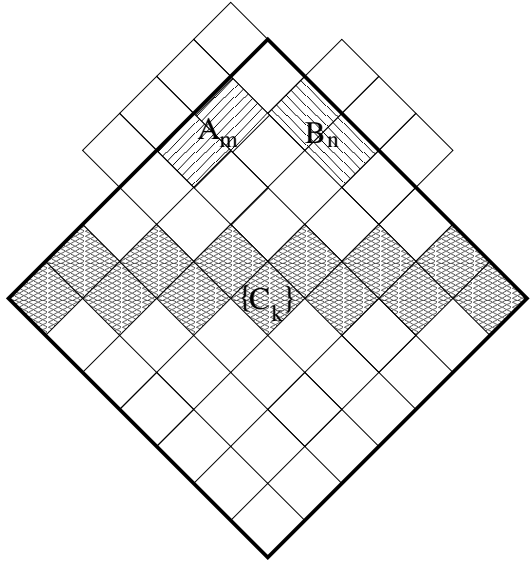}}}
\caption{Localization of a weak common cause for the correlations $\{(A_m, B_n)\}$.}
\label{strong_Bell3}
\end{figure}

First, observe that without specifying the dynamics we only can talk about \textit{weak} (commuting or noncommuting) common cause or joint common cause. The reason for this is the following. To be able to talk about the correlation between $A_m$ and $B_n$ we need to consider a local algebra containing both $A_m$ and $B_n$. To be able to talk about a common cause system $\{ C_k\}$ of this correlation we need a local algebra that contains $\{ C_k\}$ too. Let us consider the piece $V_w$ of a Cauchy surface contained in the \textit{weak} past of $\DK_A$ and $\DK_B$, which is the shadowed region in Fig. \ref{strong_Bell3}, that is the support of a hypothetical weak common cause system $\{ C_k\}$ of the correlation.
Due to primitive causality the algebra corresponding to $V_w$ is equal to the algebra $\OA(\DK_w)$, where $\DK_w:=V_w''$ is the smallest double cone containing $V_w$. Since $\DK_w$ contains both $\DK_A$ and $\DK_B$ isotony implies that $\OA(\DK_A)\vee\OA(\DK_B)\subseteq\OA(\DK_w)$ for any causal dynamics. Hence, the hypothetical weak common cause system $\{ C_k\}$ and the events $A,B$ are in the common algebra $\OA(\DK_w)$. In (Hofer-Szab\'o, Vecserny\'es 2012a) it was shown that a weak noncommuting common cause can be found for a \textit{single} correlation $\{(A, B)\}$ in local UHF-type quantum theories without the explicit knowledge of the dynamics.

\begin{figure}[ht]
\centerline{\resizebox{7cm}{!}{\includegraphics{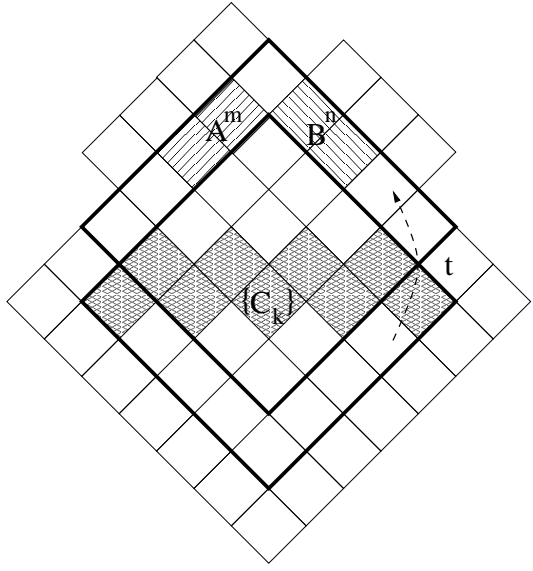}}}
\caption{Localization of a common cause for the correlations $\{(A_m, B_n)\}$.}
\label{strong_Bell3b}
\end{figure}

On the other hand, if $V_c$ is the piece of a Cauchy surface contained in the \textit{common} past of $\DK_A$ and $\DK_B$ then neither $\DK_A$ nor $\DK_B$ are contained in $\DK_c:=V_c''$. Therefore  $\OA(\DK_A)\vee\OA(\DK_B)$ and $\OA(\DK_c)$ are not related by isotony, and one should find a proper Cauchy surface piece $V_c$ in the common past and an integer time shift $t>0$ such that $\DK_A\vee\DK_B\subset V_c''+(t,0)=:\DK_c+(t,0)$ holds. (See Fig. \ref{strong_Bell3b}.)
Then the corresponding algebras are related by the dynamics
\begin{equation}
\OA(\DK_A)\vee\OA(\DK_B)\subseteq\OA(\DK_c+(t,0))=\beta^t(\OA(\DK_c)),
\end{equation}
where the last equality is due to time translation covariance. Thus for finding a joint common cause in the local quantum Ising model one has to specify the dynamics. Let us therefore choose a special set of dynamics by fixing two parameters of the causal dynamics: $\theta_2=0,\eta_2=1$. By this choice the time translation automorphism (\ref{causal_automorph2}) reads as follows:
\begin{eqnarray}\label{speccausal_automorph2}
\beta(U_{x+\fel}) &=& U_xU_{x+\fel}U_{x+1}
\end{eqnarray}
for every $x\in\IN$. 
This specification of the dynamics converts events $A_m$ and $B_n$ in (\ref{A_m})-(\ref{B_n}) into
\begin{eqnarray}
A_m &=&\fel\big(\UN+a^m_1 U_{-1} + a^m_2 U_{-1} U_{-\fel} U_0 + a^m_3 iU_{-\fel} U_0 \big) \label{A_m'} \\
B_n &=&\fel\big(\UN+b^n_1 U_1 - b^n_2 U_0 U_{\fel} U_{1}  + b^n_3 iU_0 U_{\fel} \big) \label{B_n'} 
\end{eqnarray}
and the density matrix (\ref{rho3}) into
\begin{eqnarray} \label{rho4} 
\rho &=&  \UN + \lambda \big( U_{-1} U_{-\frac{1}{2}} U_{\frac{1}{2}} U_1 - U_{-1} U_1 + U_{-\frac{1}{2}} U_{\frac{1}{2}} \big) .
\end{eqnarray}

\begin{figure}[ht]
\centerline{\resizebox{5cm}{!}{\includegraphics{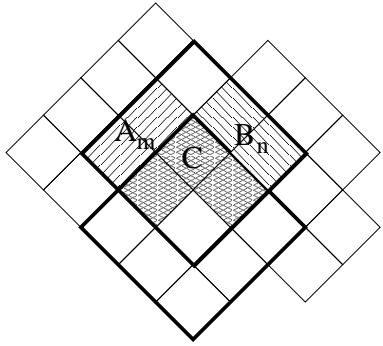}}}
\caption{Localization of the constructed common cause for the correlations $\{(A_m, B_n)\}$.}
\label{strong_Bell4}
\end{figure}

Since $\DK_c:=\DK_A\vee\DK_B-(1,0)$ is in the common past of $\DK_A$ and $\DK_B$, it would be a possible choice for the support of a hypothetical joint common cause system. However, as it will turn out soon, using this dynamics we will be able to localize a (noncommuting) common cause $\{ C,C^\perp\}$  of the correlation in a smaller spacetime region within $\DK_c$, namely in $\DK_C:= \DK_{-\fel}\vee\DK_\fel\in\SDK^m_{CS}$. (See Fig. \ref{strong_Bell4}.) 
The common cause $\{C, C^\perp\}$ localized in $\DK_C$ will be the common cause system of all the four correlations $\{(A_m, B_n); m,n=1,2\}$, i.e. $\{C, C^\perp\}$ serves as a joint common cause. 

To see the concrete form of $\{C, C^{\perp}\}$ recall that any non-central projection of rank 2 in $\OA(\DK_C)$ can be written as 
\begin{eqnarray}\label{C} 
C \equiv C({\bf c},{\bf \tilde{c}}) &=& \frac{1}{4}\left(\UN + U_{-\fel} U_{\fel}\right)\left(\UN+c_1 U_0+c_2 U_{\fel} + c_3 iU_0 U_{\fel}\right) \nonumber \\
& & + \frac{1}{4}\left(\UN - U_{-\fel} U_{\fel}\right)\left(\UN+\tilde{c}_1 U_0+\tilde{c}_2 U_{\fel} + \tilde{c}_3 iU_0 U_{\fel}\right)
\end{eqnarray}
where ${\bf c} = (c_1, c_2, c_3)$ and ${\bf \tilde{c}} = (\tilde{c}_1, \tilde{c}_2, \tilde{c}_3)$ are unit vectors in $\RE^3$. With these notations in hand we have the following
\begin{Prop}\label{NCCCCS}
Let $A_m\equiv A({\bf a^m})\in\OA(\DK_A), B_n\equiv B({\bf b^n})\in\OA(\DK_B); m,n=1,2$ be four projections defined in (\ref{A_m'})-(\ref{B_n'}) where 
${\bf a^m}$ and ${\bf b^n}$ are non-orthogonal unit vectors in $\RE^3$ establishing four correlations $\{(A_m, B_n); m,n=1,2\}$ in the state defined by the density matrix in (\ref{rho4}). Then $\{C, C^{\perp}\}$ given in (\ref{C}) is a noncommuting joint common cause of the correlations $\{(A_m, B_n)\}$ localized in $\DK_C$ if $a^m_3b^n_3 = 0$ for any $m,n=1,2$ and $c_2 =0$.
\end{Prop}

\noindent\textit{Proof.} $C_1 \equiv C({\bf c},{\bf \tilde{c}})$ and $C_2 \equiv C^{\perp} = C(-{\bf c},-{\bf \tilde{c}})$ are localized in the common past of the events $A_m$ and $B_n$. So if they satisfy the criterium
\begin{eqnarray}\label{ncqccsrew'}
(\phi\circ E)(A_mB_nC_k))(\phi\circ E)(A_m^{\perp}B_n^{\perp}C_k) &=& (\phi\circ E)(A_mB_n^{\perp}C_k)(\phi\circ E)(A_m^{\perp}B_nC_k)
\end{eqnarray}
for any $k,m,n =1,2$ then they qualify as a noncommuting joint common cause of the correlations $\{(A_m, B_n)\}$. (\ref{ncqccsrew'}) can be rewritten in the equivalent form: 
\begin{eqnarray}\label{ncqccsrew''}
Tr(\rho_k A_mB_n)\, Tr(\rho_k A_m^{\perp}B_n^{\perp}) &=& Tr(\rho_k A_mB_n^{\perp})\, Tr(\rho_k A_m^{\perp}B_n),
\end{eqnarray}
where $\rho_k:=2 C_k \rho C_k$. Using (\ref{rho4}) and (\ref{C}) one can easily calculate $\rho_1$:
\begin{eqnarray}\label{rho_c} 
\rho_1 &=& \UN + \lambda  U_{-\fel} U_\fel \nonumber \\
&& + \frac{1+\lambda}{2} c_1 (U_{-\fel} + U_\fel) 
   + \frac{1-\lambda}{2} \tilde{c}_1 (U_{-\fel} - U_\fel) \nonumber \\
&& + \frac{1+\lambda}{2} c_2 (U_0 - U_{-\fel} U_0 U_\fel) 
   - \lambda c_2 (U_{-1}U_0U_1 + U_{-1} U_{-\fel} U_0 U_\fel U_1) 
   + \frac{1-\lambda}{2} \tilde{c}_2 (U_0 + U_{-\fel} U_0 U_\fel) \nonumber \\
&& + \frac{1+\lambda}{2} c_3 i(U_{-\fel} U_0 - U_0 U_\fel) 
   + \frac{1-\lambda}{2} \tilde{c}_3 i(U_{-\fel} U_0 + U_0 U_\fel) \nonumber \\
&& + \lambda c_1 c_2  (U_{-1} U_{-\fel} U_0 U_1 + U_{-1} U_0 U_\fel U_1) \nonumber \\
&& + \lambda c^2_2  (-U_{-1} U_1 + U_{-1} U_{-\fel} U_\fel U_1) \nonumber \\
&& + \lambda c_2 c_3 i(U_{-1} U_{-\fel} U_1 - U_{-1} U_\fel U_1). 
\end{eqnarray}
$\rho_2$ can be readily obtained by substituting $(-{\bf c},-{\bf \tilde{c}})$ for $({\bf c},{\bf \tilde{c}})$. For the choice $\lambda =1$ we obtain the density matrices $\rho^s_k$ derived from the singlet state.

Substituting (\ref{A_m'}), (\ref{B_n'}) and (\ref{rho_c}) into (\ref{ncqccsrew''}) we arrive at the following equation:
\begin{eqnarray}
4[Tr(\rho_k A_mB_n)\, Tr(\rho_k A_m^\perp B_n^\perp) - 
  Tr(\rho_k A_mB_n^\perp)\, Tr(\rho_k A_m^\perp B_n)] = \nonumber \\
\lambda c_1 c_2 \, (-a^m_1 b^n_2 + a^m_2 b^n_1) - \lambda c^2_2 \, (a^m_1 b^n_1 + a^m_2 b^n_2) - \left(\lambda -\frac{(1+\lambda)^2}{4}c^2_3 
  +\frac{(1-\lambda)^2}{4}\tilde{c}^2_3 \right)  \, a^m_3 b^n_3
\end{eqnarray}
for $k=1,2$; which is zero (among others) if $a^m_3b^n_3 = 0$ for any $m,n=1,2$ and $c_2 =0$. Hence, $\{C= C({\bf c},{\bf \tilde{c}}), C^{\perp}= C(-{\bf c},-{\bf \tilde{c}})\}$ in (\ref{C}) with $c_2 =0$ is a noncommuting joint common cause localized in $\DK_C$ of the correlations $\{(A_m, B_n); m,n=1,2\}$ with $a^m_3b^n_3 = 0$.\qed
\vspace{0.2in}

\noindent
Observe that for the directions ${\bf a^m}$ and ${\bf b^n}$ defined in (\ref{a1})-(\ref{b2}) the requirement $a^m_3b^n_3 = 0$ holds for any $m,n=1,2$, hence the correlations (maximally) violating the CH inequality \textit{do} have a joint common cause---any $C$ of form (\ref{C}) with $c_2 =0$.

Finally, we show that there exists no commuting joint common cause for these correlations even without any restriction to their localization.
\begin{Prop}\label{nocommjcc}
Let $A_m\in\OA(\DK_A), B_n\in\OA(\DK_B); m,n=1,2$ be projections defined in (\ref{A_m'})-(\ref{B_n'}) with ${\bf a^m}$ and ${\bf b^n}$ given in (\ref{a1})-(\ref{b2}). 
The correlations $\{(A_m, B_n); m,n=1,2\}$ in the state (\ref{rho4}) do \textit{not} have a \textit{commuting} joint common cause $\{C_1, C_2\}$ in $\OA$.
\end{Prop}

\noindent\textit{Proof.} Since two noncommuting projections $P_1,P_2$ in $M_2(\CO)$ already generate $M_2(\CO)$, $\langle P_1, P_2\rangle=M_2(\CO)$, we have
\begin{eqnarray}
M_2(\CO)\simeq\OA(\DK_A) &=& \langle A_1, A_2\rangle=\langle U_{-1},\beta(U_{-\fel})\rangle
  = \langle U_{-1}, U_{-1}U_{-\fel}U_0\rangle, \label{Agen} \\
M_2(\CO)\simeq\OA(\DK_B) &=& \langle B_1, B_2\rangle=\langle\beta(U_\fel), U_1\rangle
  = \langle U_0U_\fel U_1, U_1\rangle.         \label{Bgen} 
\end{eqnarray} 
Hence, a commuting joint common cause $\{ C_1,C_2\}$ should be in $\OA(\DK_A)'\cap\OA(\DK_B)'\cap\OA$. Since not only $A_m, B_n$ but also $\rho$ is in $\OA(\DK_A)\vee\OA(\DK_B)$ only those $U$-monoms in $C_k$ give non-zero contribution to $Tr(\rho A_mB_nC_k)$ (and to the similar expressions containing $A_m^\perp$ or $B_m^\perp$) that are also in $\OA(\DK_A)\vee\OA(\DK_B)\simeq M_4(\CO)$. However, it is a simple matrix algebra therefore $\OA(\DK_A)'\cap\OA(\DK_B)'\cap(\OA(\DK_A)\vee\OA(\DK_B))=\CO\,\UN$. Denoting the coefficient of $\UN$ in $C_k$ by $d_k$, which is the dimension $d_k=Tr(C_k)\in(0,1)$ of the projection $C_k$, one obtains that the correlations containing $C_k$ are the original correlations (\ref{corr'}) multiplied by $d_k^2$:  
\begin{equation}
Tr(\rho A_mB_nC_k)\, Tr(\rho A_m^{\perp}B_n^{\perp}C_k) - 
Tr(\rho A_mB_n^{\perp}C_k)\, Tr(\rho A_m^{\perp}B_nC_k) 
= -\frac{\lambda}{4} \langle {\bf a^m}, {\bf b^n}\rangle d_k^2,
\end{equation}
which is non-zero for all ${\bf a^m}$ and ${\bf b^n}$ defined in (\ref{a1})-(\ref{b2}) (if $\lambda \neq 0$). Therefore $\{ C_1,C_2\}$ is not a common cause of any of the correlations $\{(A_m, B_n); m,n=1,2\}$.\qed

\section{Conclusions}

In this paper it was shown that a set of correlations between spacelike separated projections maximally violating the Clauser--Horne inequality can be given a joint common causal explanation if common causes \textit{not} commuting with the correlating events are allowed. Moreover, this noncommuting common cause could be located in the common past of the correlating events.
\vspace{0.2in}

\noindent
{\bf Acknowledgements.} This work has been supported by the Hungarian Scientific Research Fund, OTKA K-68195 and by the Fulbright Research Grant while G. H-Sz. was a Visiting Fellow at the Center for Philosophy of Science in the University of Pittsburgh.

\section*{References} 
\begin{list}
{ }{\setlength{\itemindent}{-15pt}
\setlength{\leftmargin}{15pt}}

\item J. F. Clauser, M.A. Horne, A. Shimony and R. A. Holt, ''Proposed experiment to test local hidden-variable theories,'' \textit{Phys. Rev. Lett.}, \textbf{23}, 880-884.

\item J. F. Clauser and M. A. Horne, ''Experimental consequences of objective local theories,'' \textit{Phys. Rev. D}, \textbf{10}, 526-535.

\item R. Haag, {\it Local Quantum Physics}, (Springer Verlag, Berlin, 1992). 

\item H. Halvorson and R. Clifton, ''Generic Bell correlation between arbitrary local algebras in quantum field theory,''
\textit{J. Math. Phys.}, \textbf{41}, 1711-1717 (2000).

\item G. Hofer-Szab\'o and P. Vecserny\'es ''Reichenbach's Common Cause Principle in
algebraic quantum field theory with locally finite degrees of freedom,''
\textit{Found. Phys.}, \textbf{42}, 241-255 (2012a).

\item G. Hofer-Szab\'o and P. Vecserny\'es, ''Noncommutative Common Cause Principles in algebraic quantum field theory,'' \textit{Found. Phys.}, (submitted) (2012b).

\item R. Horodecki, P. Horodecki, M. Horodecki and K. Horodecki, ''Quantum entanglement,'' \textit{Reviews of Modern Physics}, \textbf{81}, 865-942, (2009).

\item V.F. M\"uller and P. Vecserny\'es, ''The phase structure of $G$-spin models'', \textit{to be published}

\item F. Nill and K. Szlach\'anyi, ''Quantum chains of Hopf algebras with quantum double cosymmetry'' \textit{Commun. Math. Phys.}, \textbf{187} 159-200 (1997).

\item M. R\'edei, ''Reichenbach's Common Cause Principle and quantum field theory,'' \textit{Found. Phys.}, \textbf{27}, 1309--1321 (1997).

\item M. R\'edei, {\it Quantum Logic in Algebraic Approach}, (Kluwer Academic Publishers, Dordrecht, 1998).

\item M. R\'edei and J. S. Summers, ''Local primitive causality and the Common Cause Principle in quantum field theory,'' \textit{Found. Phys.}, \textbf{32}, 335-355 (2002).

\item M. R\'edei and J. S. Summers, ''Remarks on Causality in relativistic quantum field theory,'' \textit{Int. J. Theor. Phys.}, \textbf{46}, 2053–2062 (2007).

\item H. Reichenbach, {\it The Direction of Time}, (University of California Press, Los Angeles, 1956).

\item A. Shimony, ''Bell Theorems,'' \textit{Stanford Encyclopedia of Philosophy}, URL= http://plato.stanford.edu /entries/bell-theorem.

\item S. J. Summers and R. Werner, ''Maximal violation of Bell's inequalities for algebras of observables in tangent spacetime regions,'' \textit{Ann. Inst. Henri Poincar\'e -- Phys. Th\'eor.}, \textbf{49}, 215-243 (1988).

\item K. Szlach\'anyi and P. Vecserny\'es, ''Quantum symmetry and braid group statistics in $G$-spin models'' \textit{Commun. Math. Phys.}, \textbf{156}, 127-168 (1993). 

\end{list}

\end{document}